\documentclass[useAMS,usenatbib]{mn2e}
\usepackage{epsfig}

\def\gsim{\;\rlap{\lower 2.5pt
 \hbox{$\sim$}}\raise 1.5pt\hbox{$>$}\;}
\def\lsim{\;\rlap{\lower 2.5pt
   \hbox{$\sim$}}\raise 1.5pt\hbox{$<$}\;}

\def\eg{{\it e.g. }}

\def\vti{\vec{\theta}^I}
\def\vts{\vec{\theta}^S}
\def\vto{\vec{\theta}^O}
\def\vt{\vec{\theta}}

\title[Measuring the Cosmic Shear in Fourier Space]{Measuring the Cosmic Shear in Fourier Space}

\author[Jun Zhang]
{Jun Zhang\thanks{E-mail:jzhang@astro.berkeley.edu}\\ 
\\
Department of Astronomy, University of California, Berkeley, CA 94720, USA\\
}

\begin{document}


\pagerange{\pageref{firstpage}--\pageref{lastpage}} \pubyear{2006}

\maketitle

\label{firstpage}
                                      
\begin{abstract}

We propose to measure the weak cosmic shear using the spatial derivatives of the galaxy surface brightness field. The measurement should be carried out in Fourier space, in which the point spread function (PSF) can be transformed to a desired form with multiplications, and the spatial derivatives can be easily measured. This method is mathematically well defined regardless of the galaxy morphology and the form of the PSF, and involves simple procedures of image processing. Furthermore, with high resolution galaxy images, this approach allows one to probe the shape distortions of galaxy substructures, which can potentially provide much more independent shear measurements than the ellipticities of the whole galaxy. We demonstrate the efficiency of this method using computer-generated mock galaxy images.

\end{abstract}


\section{Introduction}
\label{intro}

The coherent distortions of background galaxy images by the intervening metric perturbations provide us a direct probe of the large scale mass distribution (see reviews by \citealt{bs01,wittman02,refregier03}). Recently, several groups have claimed positive detections of the weak lensing effect and obtained useful constraints on the cosmological model (\citealt{bre00,kwl00,vw00,wittman00,maoli01,rhodes01,vw01,hyg02,rrg02,bmre03,brown03,hamana03,jarvis03,rhodes04,heymans05,mbre05,vmh05,dahle06,hetterscheidt06,hoekstra06,jjbd06,schrabback06,semboloni06}). In future weak lensing observations (\eg VST-KIDS, DES\footnote{see www.darkenergysurvey.org}, VISTA darkCAM, Pan-STARRS\footnote{see pan-starrs.ifa.hawaii.edu}, LSST\footnote{see www.lsst.org}, DUNE, SNAP\footnote{see snap.lbl.gov}, JDEM\footnote{see destiny.asu.edu}), if the photometric redshift can be well calibrated, we will be able to study the dark energy properties (its abundance and equation of state) using the redshift dependence of the shear fields (\citealt{hu02,abazajian03,jain03,bernstein04,hu04,song04,takada04,takadawhite04,ishak05,simpson05,zhang05,hannestad06,schimd06,taylor06,zhan06}). By constraining the growth factor of the mass perturbation and the geometrical distance as functions of redshift separately, weak lensing provides a consistency check of the cosmological model (\citealt{kratochvil04,simpson05,zhang05,knox06}), and opens a window for testing alternative gravity theories (\citealt{acquaviva04,song05,ishak06}).  

An important and challenging job in weak lensing is to measure the weak cosmic shear (of order a few percent) from the shapes (or ellipticities) of the background galaxy images, which have large intrinsic variations. The existing methods are all based on convoluting the galaxy images with some weighting functions, and are called the INTEGRAL methods hereafter (see \citealt{tyson90,bonnet95,kaiser95,luppino97,hoekstra98,rhodes00,kaiser00,bridle01,bernstein02,refregierbacon03,massey05,kuijken06,nakajima06}). The INTEGRAL methods typically have disadvantages in three aspects: {\bf 1.} since the galaxy images are smeared by the PSF (either instrumental or environmental), the INTEGRAL methods involve at least two folds of convolutions, the math of which is complicated; {\bf 2.} the details of the methods are often sensitive to the galaxy morphology and the form of the PSF; {\bf 3.} the shear information from the shape distortions of galaxy substructures is not considered. Strictly speaking, the shapelets method (see, \eg ,Refregier 2003) may not be called an INTEGRAL method, because the galaxy weighting functions form a complete set of orthonormal shapelets which have very convenient mathematical properties. It also has the potential of measuring the cosmic shears on galaxy substructures. However, since this method requires calibrations of the intrinsic distributions of the shapelet coefficients, it has strong dependence on the galaxy morphology.
 
In this paper, we propose to use the spatial derivatives of the galaxy surface brightness field to measure the cosmic shear. This method was first used by Seljak and Zaldarriaga (1999) on CMB lensing. We generalize their analysis by including the PSF and carrying out the measurement in Fourier space. This approach is well defined regardless of the galaxy morphology and the form of the PSF, and involves simple image processing procedures. Given a high image resolution, the method can potentially probe the cosmic shear from galaxy substructures, greatly suppressing the shape noise. 

We begin by introducing the method in \S\ref{method}. In \S\ref{test}, this approach is shown to work well on different types of computer-generated mock galaxy images with general forms of PSF. A brief summary is given in \S\ref{summary}.

\section{The Method}
\label{method}

We derive the relation between the cosmic shear and the spatial derivatives of the galaxy surface brightness field without a PSF in \S\ref{withoutPSF}. In the presence of an isotropic Gaussian PSF, the relation is modified and shown in \S\ref{withisopsf}. In \S\ref{Fourier}, Fourier transformation is introduced not only to simplify the measurement of the spatial derivatives, but also to deal with general forms of PSF.   

\subsection{Without the PSF}
\label{withoutPSF}

The surface brightness on the image plane $f_I(\vti)$ and on the source plane $f_S(\vts)$ ($\vti$ and $\vts$ are the position angles on the image and source plane respectively) are related through a simple relation:
\begin{eqnarray}
\label{fifstits} 
&&f_I(\vti)=f_S(\vts)\\ \nonumber
&&\vti=\mathbf{A}\vts
\end{eqnarray}
where $\mathbf{A}_{ij}=\delta_{ij}+\Phi_{ij}$, and $\Phi_{ij}=\partial\delta\theta^I_i/\partial\theta^S_j$ are the spatial derivatives of the lensing deflection angle, which can be expressed in terms of the convergence $\kappa=(\Phi_{xx}+\Phi_{yy})/2$ and the two shear components $\gamma_1=(\Phi_{xx}-\Phi_{yy})/2$ and $\gamma_2=\Phi_{xy}$. Using eq.[\ref{fifstits}], we get:
\begin{eqnarray}
\label{dfidfs}
\frac{\partial f_I}{\partial \theta^I_i}&=&\frac{\partial\theta^S_j}{\partial\theta^I_i}\frac{\partial f_S}{\partial\theta^S_j}\\ \nonumber
&=&(\delta_{ij}-\Phi_{ij})\frac{\partial f_S}{\partial\theta^S_j}
\end{eqnarray}
where we have implicitly assumed that $\Phi_{ij}$ is small, which is true for weak lensing. Assuming the original surface brightness field $f_S$ is isotropic on the source plane, the quadratic combinations of the derivatives of the lensed image provide a direct measure of the shear components (\citealt{seljak99}):
\begin{eqnarray}   
\label{shear12}
&&\frac{1}{2}\frac{\langle (\partial_xf_I)^2-(\partial_yf_I)^2\rangle}{\langle (\partial_xf_I)^2+(\partial_yf_I)^2\rangle}=-\gamma_1 \\ \nonumber
&&\frac{\langle\partial_xf_I\partial_yf_I\rangle}{\langle (\partial_xf_I)^2+(\partial_yf_I)^2\rangle}=-\gamma_2
\end{eqnarray}
where the averages are taken over the whole galaxy.

\subsection{With an Isotropic Gaussian PSF}
\label{withisopsf}

The presence of PSF brings both advantages and disadvantages. On the positive side, the PSF smooths out the galaxy surface brightness field, which is originally not differentiable due to structures on arbitrarily small scales. On the other hand, the convolution of the galaxy image with the PSF leads to a nontrivial modification to eq.[\ref{shear12}], the form of which is calculated in this section. For simplicity, we assume the PSF is isotropic and Gaussian. General forms of PSF will be discussed in \S\ref{Fourier}. 
  
The observed galaxy surface brightness distribution $f_O$ is related to $f_I$ via:
\begin{equation}
\label{fofi}
f_O(\vt)=\int d^2\vti W_{\beta}(\vt -\vti)f_I(\vti)
\end{equation}  
where $W_{\beta}$ is the Gaussian PSF with scale length $\beta$:
\begin{equation}
\label{wbeta}
W_{\beta}(\vt)=\frac{1}{2\pi\beta^2}\exp\left(-\frac{\vert\vt\vert^2}{2\beta^2}\right)
\end{equation}
Using eq.[\ref{fifstits}] to replace $f_I$ with $f_S$ and $\vti$ with $\vts$ in eq.[\ref{fofi}], we get:
\begin{equation}
\label{fofs}
f_O(\vt)=\vert{\rm det}(\mathbf{A})\vert\int d^2\vts W_{\beta}(\vt -\mathbf{A}\vts)f_S(\vts)
\end{equation}
or equivalently:
\begin{eqnarray}
\label{fofs2}
f_O(\mathbf{A}\vt)&=&\vert{\rm det}(\mathbf{A})\vert\int d^2\vts W_{\beta}[\mathbf{A}(\vt-\vts)]f_S(\vts)\\ \nonumber
&\doteq&\vert{\rm det}(\mathbf{A})\vert\int d^2\vts f_S(\vts)W_{\beta}(\vt-\vts)\\ \nonumber
&\times&\left[1-(\vt-\vts)\cdot(\mathbf{A}-\mathbf{I})\cdot(\vt-\vts)/\beta^2\right]
\end{eqnarray}
where $\mathbf{I}$ is the $2\times 2$ unitary matrix. Note that the second part of eq.[\ref{fofs2}] is a result of Taylor expansion of the term $W_{\beta}[\mathbf{A}(\vt-\vts)]$ due to the small amplitudes of the lensing components $\Phi_{ij}$. For convenience, let us define:
\begin{equation}
\label{F_S}
F_S(\vt)=\int d^2\vts f_S(\vts)W_{\beta}(\vt-\vts)
\end{equation}
which is the surface brightness field we would observe in absence of lensing. Eq.[\ref{fofs2}] can then be re-written as:
\begin{eqnarray}
\label{fofs3}
\frac{f_O(\mathbf{A}\vt)}{\vert{\rm det}(\mathbf{A})\vert}&=&(1-\Phi_{xx}-\Phi_{yy})F_S(\vt)\\ \nonumber
&-&\beta^2\left(\Phi_{xx}\frac{\partial^2F_S}{\partial\theta_x^2}+2\Phi_{xy}\frac{\partial^2F_S}{\partial\theta_x\partial\theta_y}+\Phi_{yy}\frac{\partial^2F_S}{\partial\theta_y^2}\right)
\end{eqnarray}
Let $\vto=\mathbf{A}\vt$, then:
\begin{equation}
\label{fvto}
\frac{\partial f_O(\vto)}{\partial\theta^O_i}=(\mathbf{A}^{-1})_{ij}\frac{\partial f_O(\mathbf{A}\vt)}{\partial\theta_j}
\end{equation}
Using eq.[\ref{fofs3}] and eq.[\ref{fvto}], it is not hard to express the derivatives of $f_O$ in terms of the derivatives of $F_S$:
\begin{eqnarray}
\label{dfodFs}
\frac{\partial_xf_O}{\vert{\rm det}(\mathbf{A})\vert}&=&(1-2\Phi_{xx}-\Phi_{yy})F_x-\Phi_{xy}F_y\\ \nonumber
&-&\beta^2(\Phi_{xx}F_{xxx}+2\Phi_{xy}F_{xxy}+\Phi_{yy}F_{xyy})\\ \nonumber
\frac{\partial_yf_O}{\vert{\rm det}(\mathbf{A})\vert}&=&(1-\Phi_{xx}-2\Phi_{yy})F_y-\Phi_{xy}F_x\\ \nonumber
&-&\beta^2(\Phi_{yy}F_{yyy}+2\Phi_{xy}F_{xyy}+\Phi_{xx}F_{xxy})
\end{eqnarray}
where 
\begin{equation}
\label{defineDF}
F_{i_1\cdot\cdot\cdot i_n}=\frac{\partial^nF_S}{\partial\theta_{i_1}\cdot\cdot\cdot\partial\theta_{i_n}}
\end{equation}
Note that we have implicitly assumed that the spatial fluctuation of the cosmic shear is negligible on galactic scales. Assuming the distribution of $F_S$ is isotropic, we obtain the following relation between the shear components and the spatial derivatives of the surface brightness field:
\begin{eqnarray}
\label{shear12PSF}
&&\frac{1}{2}\frac{\langle (\partial_xf_O)^2-(\partial_yf_O)^2\rangle}{\langle (\partial_xf_O)^2+(\partial_yf_O)^2\rangle+\Delta}=-\gamma_1 \\ \nonumber
&&\frac{\langle\partial_xf_O\partial_yf_O\rangle}{\langle (\partial_xf_O)^2+(\partial_yf_O)^2\rangle+\Delta}=-\gamma_2
\end{eqnarray}
where
\begin{equation}
\label{Delta}
\Delta=\frac{\beta^2}{2}\langle\vec{\nabla}f_O\cdot\vec{\nabla}(\nabla^2f_O)\rangle
\end{equation}

The derivation of eq.[\ref{shear12PSF}] and eq.[\ref{Delta}] is shown in the appendix. Note that in the limit when the galaxy image is very smooth over the scale length $\beta$, the correction $\Delta$ approaches zero, eq.[\ref{shear12PSF}] then reduces to eq.[\ref{shear12}]. 

\subsection{Fourier Transform and General PSF}
\label{Fourier}

For the method to become useful, there are at least two remaining issues to be addressed: {\bf 1.} how to measure the spatial derivatives of the surface brightness field; {\bf 2.} how to deal with other forms of the PSF. It turns out that Fourier transformation provides a solution to both problems.

Since convolutions in real space correspond to multiplications in Fourier space, one can easily transform the PSF to a desired form (an isotropic Gaussian form in our case) by multiplying the Fourier modes of the observed image with the ratios between the Fourier modes of the desired PSF and those of the original PSF (known from calibrations with stars). This operation is usually well defined if the scale length of the desired PSF is larger than that of the original PSF. Moreover, it turns out that for the purpose of measuring the cosmic shear, one does not need to transform the new image back to real space, because the derivatives of the surface brightness field can be more easily measured in Fourier space.

As an example, we show how to measure quantities such as $\langle\vert\vec{\nabla}f\vert^2\rangle$, where $f$ is the surface brightness field of interest. First of all, the distribution $f$ in real space should be sampled with an interval $\Delta\theta$ which is a few times less than the size of the PSF to avoid translating high frequency power into the frequency range determined by the sampling resolution through discrete Fourier transform (\citealt{press92}). In other words, the galaxy image should be ``oversampled'' to avoid aliasing power from small scales in the discrete Fourier transform. For undersampled images, one can smooth the images with an additional large enough PSF, which can be treated as a part of the PSF from the instrumentation, and therefore does not affect our discussion below. Similarly, to avoid such aliasing power at low frequency, the box size for the Fourier transform should be a few times larger than the image size. Given this setup, the Fourier transform of the image is defined as:
\begin{equation}
\label{fourierf}
\tilde{f}(l_i,l_j)=\Delta\theta^2\sum_{m=0}^{N-1}\sum_{n=0}^{N-1}f(\theta_m,\theta_n)\exp\left[i(\theta_ml_i+\theta_nl_j)\right]
\end{equation}
where 
\begin{eqnarray}
\label{thetal}
&&\theta_{m(n)}=m(n)\times\Delta\theta, \mbox{   } m(n)=0,1,...,N-1\\ \nonumber
&&l_{i(j)}=i(j)\times\Delta l, \mbox{   } i(j)=-N/2,..,N/2 \\ \nonumber
&&\Delta l=2\pi/(N\Delta\theta)
\end{eqnarray}
N is the box size, chosen to be a power of $2$ for the Fast Fourier Transform. It is now straightforward to show that $\langle\vert\vec{\nabla}f\vert^2\rangle$ can be expressed as the sum over the Fourier modes weighted by the wave numbers:
\begin{eqnarray}
\label{fourierfinverse}
&&\sum_{m=0}^{N-1}\sum_{n=0}^{N-1}\vert\vec{\nabla}f(\theta_m,\theta_n)\vert^2\\ \nonumber
&=&\frac{1}{N^2\Delta\theta^4}\sum_{i=-N/2}^{N/2}\sum_{j=-N/2}^{N/2}\vert\tilde{f}(l_i,l_j)\vert^2(l_i^2+l_j^2)
\end{eqnarray}
Eq.[\ref{fourierfinverse}] gives exactly the quantity $\langle\vert\vec{\nabla}f\vert^2\rangle$ multiplied by the number of bright pixels covered by the galaxy, because the dark pixels have no contributions\footnote{In the presence of noise, extra procedures may be required to clean the galaxy map before the Fourier transform. We shall discuss this in a future paper.}. Similarly, we can calculate the other terms in eq.[\ref{shear12PSF}] in Fourier space. Note that for the purpose of obtaining $\gamma_1$ and $\gamma_2$, it is not necessary to calculate the number of bright pixels because it appears in both the nominator and the denominator in eq.[\ref{shear12PSF}].

\section{The Test}
\label{test}

This section is organized as follows: in \S\ref{diskgalaxy}, we test the method using mock regular galaxies smeared by different forms of PSF; in \S\ref{irregulargalaxy}, using mock irregular galaxies generated by 2-D random walks, we further demonstrate the usefulness of this approach on galaxies with a different morphology, and explore the possibility of suppressing the shape noise in the shear measurements by including the information from galaxy substructures. 

\subsection{With Mock Regular Galaxies}
\label{diskgalaxy}

Each regular galaxy in our simulation contains a thin circular disk with an exponential profile and a co-axial de Vaucouleurs-type spheroidal component (\citealt{vaucouleurs91}). When viewed face-on, the surface brightness distribution (before lensing and smearing by the PSF) of the galaxy can be parameterized as:
\begin{equation}
\label{diskprofile}
f(r)=\exp(-r/r_d)+f_{s/d}\exp\left[-(r/r_s)^{\frac{1}{4}}\right]
\end{equation}
where $r$ is the distance to the galaxy center, $r_d$($r_s$) is the scale length of the disk(spheroid), and $f_{s/d}$ determines the relative importance of the spheroid. The overall luminosity of the galaxy is only important in the presence of noise, which will be discussed in a future paper. 

Our simulation box is $128\times128$. We choose $r_d$ to be $1/32$ of the box size of the simulation, $r_s=r_d/2$, and $f_{s/d}=1$. Note that changing these particular numbers does not affect our main conclusions. Once the galaxy's face-on image is generated, it is projected onto the source plane with a random inclination angle along a random direction perpendicular to the line of sight\footnote{The intrinsic flattening parameter $q$ of the spheroid part is set to one for simplicity.}. 

\begin{figure}
\setlength{\epsfxsize}{0.5\textwidth}
\setlength{\epsfysize}{0.25\textwidth}
\centerline{\epsfbox{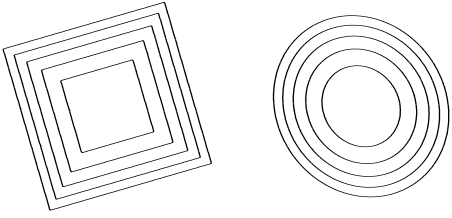}}
\caption{Two PSF models used in the simulations of \S\ref{diskgalaxy}. The contours mark 3.2\%, 6.3\%, 12.5\%, 25\%, and 50\% of the peak intensity. The left and right one correspond to $W_r^{(1)}$ and $W_r^{(2)}$ (rotated by certain angles) in eq.[\ref{PSFforms}] respectively. 
}
\label{PSFs}
\end{figure}

The projected galaxy image is subsequently distorted by a constant cosmic shear and smeared by the PSF in real space. We consider two PSF models given by the following forms rotated by certain angles (shown in Fig.\ref{PSFs}):
\begin{eqnarray}
\label{PSFforms}
&&W_r^{(1)}(x,y)\propto\exp\left[-(\vert x-y\vert +\vert x+y\vert)^2/(8r^2)\right]\\ \nonumber
&&W_r^{(2)}(x,y)\propto\exp\left[-(x^2+0.8y^2)/(2r^2)\right]\\ \nonumber
\end{eqnarray}
where $r$ is the scale length, which is equal to six times the grid size, comparable to the galaxy size. The shear components $(\gamma_1,\gamma_2)$ are chosen to be $(-0.012,0.035)$, $(-0.032,-0.005)$, $(0.01,0.02)$ for $W_r^{(1)}$, and $(0.015,-0.024)$, $(0.05,0.01)$, $(-0.04,-0.04)$ for $W_r^{(2)}$. 

\begin{figure}
\setlength{\epsfxsize}{0.5\textwidth}
\centerline{\epsfbox{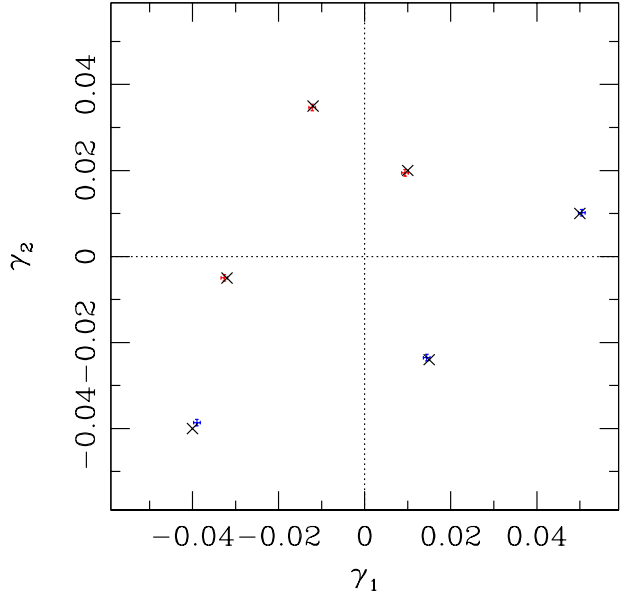}}
\caption{The measured values of $(\gamma_1,\gamma_2)$ (averaged over $10000$ mock regular galaxies) are plotted as red (from galaxies smoothed by $W_r^{(1)}$) and blue (from galaxies smoothed by $W_r^{(2)}$) dots with (small) $1\sigma$ error bars. The centers of the black crosses are the input values of $(\gamma_1,\gamma_2)$. Note that to be general, $W_r^{(1)}$ and $W_r^{(2)}$ in the simulations have been rotated by certain angles with respect to their definitions in eq.[\ref{PSFforms}].
}
\label{shear_disk}
\end{figure}

To measure the cosmic shear, we follow the procedures described in \S\ref{Fourier}. The desired PSF has an isotropic Gaussian form with a scale length about $4/3$ times that of the original PSF. The results are plotted in Fig.\ref{shear_disk}. The results are consistent within $1\sigma$ error regardless of the form of the PSF. 

\subsection{With Mock Irregular Galaxies}
\label{irregulargalaxy}

Our irregular galaxies are generated using 2-D random walks. The random walk starts from the center of the simulation box for 20000 steps, each of which is equal to the grid size of the simulation box (which is now $1024\times1024$). Once the distance from the center is more than $1/6$ of the box size, the walk starts from the center again to finish the rest of the steps. The surface brightness of the galaxy is equal to the density of the trajectories. Note that these galaxies naturally have abundant substructures, which are useful not only for further testing the method, but also for illustrating how much lensing information may be contained in the substructures. We caution that our random-walk-type galaxies are not based on any physical models, therefore they do not necessarily mimic observed irregular galaxies. In a future paper, more realistic galaxy models will be adopted to study this topic. 

\begin{figure}
\setlength{\epsfxsize}{0.5\textwidth}
\centerline{\epsfbox{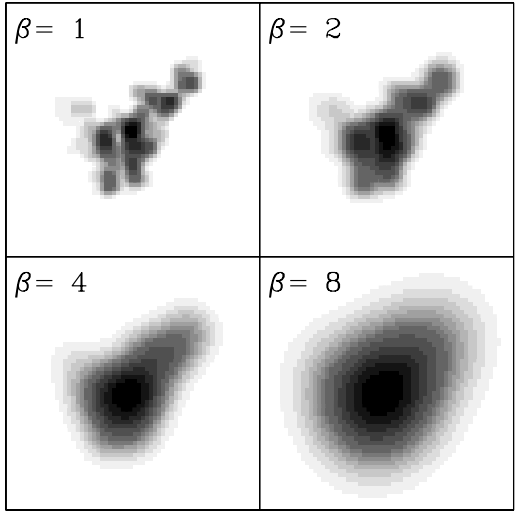}}
\caption{Sample images showing our mock irregular galaxies generated using 2-D random walks under four different angular resolutions.}
\label{rwgalaxy}
\end{figure}

For the purpose of this section, we smooth the galaxies directly with the isotropic Gaussian PSF of different scale lengths, which correspond to different angular resolutions. The scale length $\beta$ (defined in eq.[\ref{wbeta}]) is chosen to be $1/256$, $1/128$, $1/64$, and $1/32$ of the box size (roughly corresponding to $1/85$, $1/43$, $1/21$, and $1/10$ of the galaxy size). Fig.\ref{rwgalaxy} shows typical images of our irregular galaxy under these four different angular resolutions. For convenience, we plot the minimum $\beta$ as unity in the figures of this section.

\begin{figure}
\setlength{\epsfxsize}{0.5\textwidth}
\centerline{\epsfbox{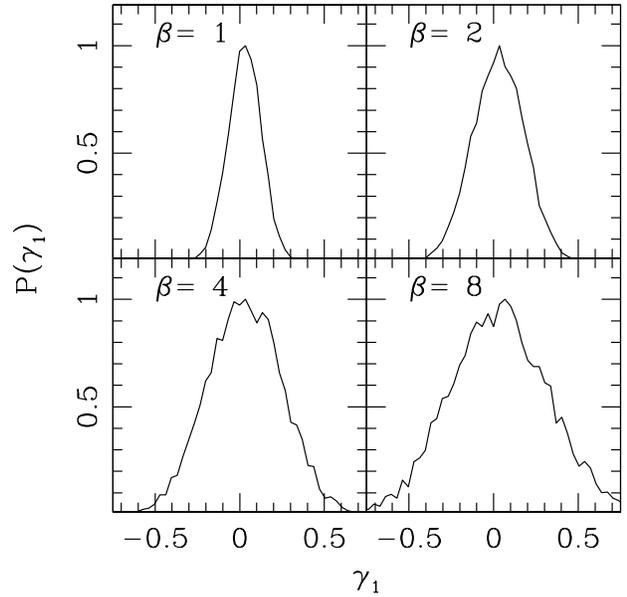}}
\caption{The probability distribution functions (PDF) of the measured $\gamma_1$ for four different image resolutions. The plots are all generated using $10000$ mock irregular galaxies. The peak of the PDF is fixed at $1$ in each plot.}
\label{results}
\end{figure}

\begin{figure}
\setlength{\epsfxsize}{0.5\textwidth}
\centerline{\epsfbox{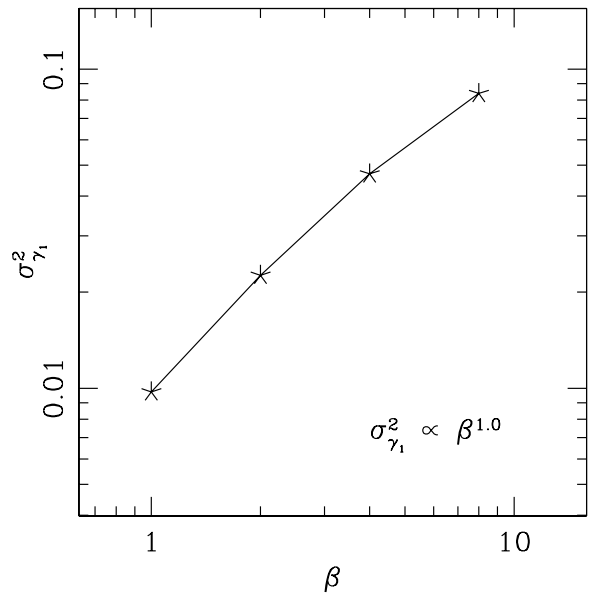}}
\caption{The variance of $\gamma_1$ for four different values of $\beta$.
}
\label{sigma}
\end{figure}

The shear component $\gamma_2$ is set to zero, and $\gamma_1$ is fixed at $0.03$. After averaging over $10000$ irregular galaxies, we find that the measured $\gamma_1$ is $0.0324\pm0.0029$ for $\beta=8$, $0.0301\pm0.0022$ for $\beta=4$, $0.0291\pm0.0015$ for $\beta=2$, and $0.0293\pm0.0010$ for $\beta=1$. More interestingly, as shown in Fig.\ref{results}, the statistical error bar is found to decrease significantly when the angular resolution is increased. This is further illustrated in Fig.\ref{sigma}, which shows an approximate power-law relation between the measured variance of $\gamma_1$ and $\beta$, the exponent of which is close to one. Note that as the angular resolution increases, one gets additional information on the cosmic shears from the galaxy substructures. If we naively assume that each bright pixel on the galaxy map provides an independent measurement of the cosmic shear, we expect the variance of the measured cosmic shear to scale as the inverse of the number of the bright pixels, or $\beta^d$, where $d$ is the Hausdorff dimension of the galaxy image (\citealt{hausdorff1919}). Since the Hausdorff dimension of our random-walk-generated irregular galaxies is $2$ (\citealt{falconer86}), the variance of $\gamma_1$ should scale as $\beta^{2}$, which is not too far from what we have observed in our numerical experiments. In reality, substructures generated by the 2-D random walks are correlated at some unknown level, therefore, the observed exponent indicated in Fig.\ref{sigma} is less than the Hausdorff dimension. 

\section{Summary}
\label{summary}

We have presented a simple approach of measuring the weak cosmic shear using the spatial derivatives of the galaxy surface brightness field. The measurement should be carried out in Fourier space, in which it is easy to evaluate the spatial derivatives and to transform the PSF to a desired form. The accuracy of the method is demonstrated using computer-generated mock regular and irregular galaxies. We find no systematic errors on the measured shear components in the numerical experiments. 

Given high image resolutions, this new method may reduce the shape noise in the shear measurement significantly, because it takes into account the shape information on the galaxy substructures. Using the mock irregular galaxies generated by 2-D random walks, we have shown that the variance of the measured shear is indeed suppressed by a large factor when the image resolution is increased. This example encourages us to test this method on real galaxies of a wide range of morphology classes in a future paper by joining the Shear TEsting programme (\citealt{heymans06,massey06}), the results of which may be useful for optimizing the signal to noise ratio in shear measurements and planning future weak lensing survey.

\section*{Acknowledgements}

We thank Chung-Pei Ma, Tony Tyson, Martin White, David Wittman for useful discussions, Gary Bernstein, Lam Hui, Bhuvnesh Jain, Nick Kaiser, and the anonymous referee for comments on an earlier version of this manuscript. Research for this work is supported by NASA, and by the TAC Fellowship of UC Berkeley.

\bibliographystyle{mn2e}

\section*{Appendix -- Relating the Cosmic Shears with the Spatial Derivatives of the Surface Brightness field}
\label{appendix}

From eq.[\ref{dfodFs}], we have:
\begin{eqnarray}
\label{appendix1}
&&\frac{1}{\vert{\rm det}(\mathbf{A})\vert^2}\left[(\partial_xf_O)^2-(\partial_yf_O)^2\right]\\ \nonumber
&=&(1-6\kappa)(F_x^2-F_y^2)-2\gamma_1(F_x^2+F_y^2)\\ \nonumber
&-&\beta^2\left[2\kappa\Pi_1+\gamma_1(\Lambda+\Upsilon_1)+\gamma_2(\Upsilon_2-\tilde{\Lambda})\right]
\end{eqnarray}
and 
\begin{eqnarray}
\label{appendix2}
&&\frac{2}{\vert{\rm det}(\mathbf{A})\vert^2}\partial_xf_O\partial_yf_O\\ \nonumber
&=&2(1-6\kappa)F_xF_y - 2\gamma_2(F_x^2+F_y^2)\\ \nonumber
&-&\beta^2\left[2\kappa\Pi_2+\gamma_1(\tilde{\Lambda}+\Upsilon_2)+\gamma_2(\Lambda-\Upsilon_1)\right]
\end{eqnarray}
where
\begin{eqnarray}
\label{appendix3}
&&\Lambda=F_xF_{xxx}+F_xF_{xyy}+F_yF_{xxy}+F_yF_{yyy}\\ \nonumber
&&\tilde{\Lambda}=F_yF_{xxx}+F_yF_{xyy}-F_xF_{xxy}-F_xF_{yyy}\\ \nonumber
&&\Pi_1=F_xF_{xxx}+F_xF_{xyy}-F_yF_{xxy}-F_yF_{yyy}\\ \nonumber
&&\Pi_2=F_xF_{xxy}+F_xF_{yyy}+F_yF_{xyy}+F_yF_{xxx}\\ \nonumber
&&\Upsilon_1=F_xF_{xxx}-3F_xF_{xyy}-3F_yF_{xxy}+F_yF_{yyy}\\ \nonumber
&&\Upsilon_2=F_yF_{xxx}-3F_yF_{xyy}+3F_xF_{xxy}-F_xF_{yyy}
\end{eqnarray}

Note that according to the definitions in eq.[\ref{appendix3}], $\Lambda$ is a scalar, $\tilde{\Lambda}$ is a pseudo scalar, $\Pi_1+i\Pi_2$ is a spin-2 field, and $\Upsilon_1+i\Upsilon_2$ is a spin-4 field. If the intrinsic surface brightness distribution is isotropic, the spatial averages of $\tilde{\Lambda}$, $\Pi_1$, $\Pi_2$, $\Upsilon_1$, and $\Upsilon_2$ must vanish. As a result of this, we have:
\begin{eqnarray}  
\label{appendix4}
&&\frac{1}{2}\langle(\partial_xf_O)^2-(\partial_yf_O)^2\rangle\\ \nonumber
&=&-\gamma_1\left(\langle F_x^2+F_y^2\rangle+\frac{\beta^2}{2}\langle\Lambda\rangle\right)\\ \nonumber
&&\langle\partial_xf_O\partial_yf_O\rangle\\ \nonumber
&=&-\gamma_2\left(\langle F_x^2+F_y^2\rangle+\frac{\beta^2}{2}\langle\Lambda\rangle\right)
\end{eqnarray}
We have neglected the factor $\vert{\rm det}(\mathbf{A})\vert$ which is equal to unity to the 0th order. Using the fact that $\Lambda=\vec{\nabla}F_S\cdot\vec{\nabla}(\nabla^2F_S)$, and $F_S=f_O$ to the 0th order, it is now straightforward to prove eq.[\ref{shear12PSF}]. 
  
\label{lastpage}

\end{document}